\documentclass[12pt,letterpaper]{article}
\usepackage[utf8]{inputenc}

\pdfoutput=1
\usepackage{color}
\definecolor{darkblue}{rgb}{0.1,0.1,.7}
\usepackage[colorlinks, linkcolor=darkblue, citecolor=darkblue, urlcolor=darkblue, linktocpage]{hyperref}
\usepackage[]{amsmath}
\usepackage[]{graphicx}
\usepackage[]{latexsym}
\usepackage[utf8]{inputenc}
\usepackage{slashed,graphicx,color,amsmath,amssymb}
\usepackage[mathscr]{eucal}
\usepackage{mathrsfs}
\usepackage[margin=10pt,font=small,labelfont=bf]{caption}
\usepackage{amscd}
\usepackage{bm}
\usepackage{xcolor}
\usepackage{upgreek}
\usepackage[square, comma, sort&compress,numbers]{natbib}
\usepackage[all,cmtip]{xy}
\usepackage[margin=1.7in]{geometry}
\usepackage{cleveref}
\usepackage[color=cyan!30!white,linecolor=red,textsize=footnotesize]{todonotes}
\usepackage{soul}
\numberwithin{equation}{section}
\hypersetup{
	pdfstartview={FitH},    
	pdftitle={Adding Flavor to the Narain ensemble},    
	pdfauthor={Datta, Duary, Kraus, Maity, Maloney},     
	colorlinks=true,       
	linkcolor=blue,          
	citecolor=blue!59!black! ,        
	filecolor=magenta,      
	urlcolor=blue           
}

\newcommand{\rt}{\mathrm{t}}

\def\btau{{\bar{\tau}}}
\def\Z{\mathbb{Z}}
\def\pd{\partial}

\def\nn{\nonumber}
\def\pd{\partial}

\def\l1{{{1-loop}}}

\def\n1{\Bigg|_{n=1}}

\def\n{{(n)}}

\def\pd{\partial}

\def\beq{\begin{equation}}
\def\eeq{\end{equation}}

\def\bea{\begin{eqnarray}}
\def\eea{\end{eqnarray}}
\def\nn{\nonumber}
\def\pd{\partial}

\def\l1{{\text{1-loop}}}

\def\n1{\Bigg|_{n=1}}

\def\n{{(n)}}

\def\vev#1{\langle{#1}\rangle}

\def\be{\begin{equation}}
\def\ee{\end{equation}}
\def\bal{\begin{array}{l}}
\def\ba#1{\begin{array}{#1}}  
	\def\ea{\end{array}}
\def\bea{\begin{eqnarray}}
\def\eea{\end{eqnarray}}
\def\beas{\begin{eqnarray*}}
	\def\eeas{\end{eqnarray*}}

\def\nn{\\\nonumber}

\def\vev#1{\langle #1 \rangle}

\def\eps{\epsilon}

\def\nn{\nonumber}
\def\bit{\begin{item}}
	\def\eit{\end{item}}
\def\benu{\begin{enumerate}}
	\def\eenu{\end{enumerate}}

\def\p{\partial}
\def\Ab{\overline{A}}
\def\Jb{\overline{J}}
\def\wb{\overline{w}}
\def\zb{\overline{z}}

\def\taub{\overline{\tau}}
\def\Lb{\overline{L}}
\def\rt{\rightarrow}
\def\taub{\overline{\tau}}

\DeclareFontFamily{U}{wncy}{}
\DeclareFontShape{U}{wncy}{m}{n}{<->wncyr10}{}
\DeclareSymbolFont{mcy}{U}{wncy}{m}{n}
\DeclareMathSymbol{\Sha}{\mathord}{mcy}{"58}

 \makeatletter
 \g@addto@macro\bfseries{\boldmath}
 \makeatother

\makeatletter
\if@todonotes@disabled

\else

\fi
\makeatother

\begin{document}

\definecolor{tinge}{RGB}{255, 244, 195}
\sethlcolor{tinge}
\setstcolor{red}

\vspace*{-.8in} \thispagestyle{empty}
\begin{flushright}
	\texttt{CERN-TH-2021-020}
\end{flushright}
\vspace{.2in} {\Large
\begin{center}
{\LARGE \bf  Adding Flavor to the Narain Ensemble}
\end{center}}
\vspace{.2in}
\begin{center}
{Shouvik Datta$^1$, Sarthak Duary$^{2}$, Per Kraus$^3$, Pronobesh Maity$^2$ \& Alexander Maloney$^4$ }
\\
\vspace{.3in}
\small{
$^1$  \textit{Department of Theoretical Physics, CERN,\\
	1 Esplanade des Particules, Geneva 23, CH-1211, Switzerland.}\\
\vspace{.2cm}
$^2$ \textit{International Centre for Theoretical Sciences-TIFR,\\
	Shivakote, Hesaraghatta Hobli, Bengaluru North 560 089, India.}
\\ \vspace{.2cm}
$^3$ \textit{Mani L. Bhaumik Institute for Theoretical Physics,\\
	Department of Physics and Astronomy,\\
	University of California, Los Angeles, CA 90095, USA.}
\\ \vspace{.2cm}
$^4$ \textit{Department of Physics, McGill University,\\ Montr\'{e}al, QC, H3A 2T8, Canada.}
\\ 
%
\vspace{.1cm}

} \vspace{0cm}


\end{center}

\vspace{.5in}

\begin{abstract}
\normalsize
We revisit the proposal that the ensemble average over free boson CFTs in two dimensions --- parameterized by Narain's moduli space ---
is dual to an exotic theory of gravity in three dimensions dubbed $U(1)$ gravity.
We consider flavored partition functions, where the usual genus $g$ partition function is weighted by Wilson lines   coupled to the conserved $U(1)$ currents of these theories.  These flavored partition functions obey a heat equation which relates deformations of the Riemann surface moduli to those of the
chemical potentials which measure these $U(1)$ charges.  This allows us to derive a Siegel-Weil formula which computes the average of these flavored partition functions.
The result takes the form of a ``sum over geometries,'' albeit with modifications relative to the unflavored case.
\end{abstract}

\vskip 1cm \hspace{0.7cm}

\newpage

\setcounter{page}{1}

\noindent\rule{\textwidth}{.1pt}\vspace{-1.2cm}
\begingroup
\hypersetup{linkcolor=black}
\tableofcontents
\endgroup
\noindent\rule{\textwidth}{.2pt}

\section{Introduction}


One of the most striking observations in the study of quantum gravity is that
certain simple gravitational theories -- primarily those in a low number of space-time dimensions --  appear to be described not by a single quantum theory, but rather by an ensemble average of many theories.  
This phenomenon was initially described for Jackiw-Teitelboim gravity in AdS$_2$, which is dual to a random matrix theory \cite{SSS}. 
This is a prototypical example of an AdS$_2$/CFT$_1$ duality.
In order to understand higher dimensional versions of this phenomenon, one would like to understand ensembles of random conformal field theories which are dual to putative theories of gravity in Anti-de Sitter space.  At first sight, constructing a random conformal field theory seems quite difficult, as it would involve an ensemble average over the space of conformal field theories, a space which is itself quite poorly understood.  For this reason, recent work in this direction \cite{Maloney:2020nni,Afkhami-Jeddi:2020ezh} has focused on CFTs with enhanced symmetry algebras where the space of CFTs can be understood precisely (related works in this direction include  \cite{Cotler:2020hgz,Cotler:2020ugk,Perez:2020klz,Meruliya:2021utr,Dymarsky:2020pzc}).

The natural starting point is perhaps the simplest possible family of two dimensional CFTs: unitary, compact CFTs with $U(1)^D \times U(1)^D$ current algebra and central charge $c=D$.  These are simply theories of $D$ free compact bosons, and the data which defines such a theory is an
even, self-dual lattice of signature $(D,D)$.  The moduli space of such theories is the homogeneous space \cite{Narain:1985jj,Narain:1986am}
\be
{\cal M}_D={ O(D,D,\Z)\backslash O(D,D) \slash O(D)\times O(D)}~.
\ee
This space has finite volume, and a unique homogeneous metric which can be used to define a probability distribution on the associated space of CFTs.  The
work of \cite{Maloney:2020nni,Afkhami-Jeddi:2020ezh} argued that this ensemble average is dual to an exotic three dimensional theory of gravity in AdS$_3$ dubbed ``$U(1)$ gravity.''  This theory of gravity includes as its perturbative degrees of freedom a $U(1)^{2D}$ Chern-Simons theory describing the gauge dynamics dual to the $U(1)^{2D}$ global symmetry in the boundary.  The non-perturbative structure of the theory is defined by a sum over geometries in the bulk.  Together these ingredients were shown to reproduce the ensemble average of the genus $g$ partition function, which was computed using the Siegel-Weil formula in terms of a real analytic Eisenstein series \cite{siegel1951indefinite,weil1964certains,weil1965formule}.

The genus $g$ partition function, however, is not the most general observable of the theory.  The theory contains global $U(1)$ charges, so one can in addition
consider ``flavored'' partition functions which include fugacities that couple to these global $U(1)$ charges.  For example, on the torus one can consider the flavored partition function
 \be
Z\!\left(\tau,\btau,z^I_L,z^I_R\right) =\,  {\rm Tr} \left[e^{2\pi i \tau (L_0-\frac{c}{24})}e^{-2\pi i (L_0-\frac{c}{24})}e^{2\pi i z_L^I J^I_0}e^{-2\pi i z_R^I \bar J^I_0}\right]~,
 \ee
 which depends on both the conformal structure parameter $\tau$ as well as a $D$-component vector $(z_L^I,z_R^I)$ of chemical potentials.  Geometrically, these chemical potentials can be interpreted as background Wilson lines which couple to the global $U(1)$ charges $(Q^I,\bar Q^I)$ of a state.  At higher genus, one can consider more general flavored partition functions which include Wilson lines wrapping arbitrary cycles in the boundary surface.

 The natural question is then: is there a version of the Siegel-Weil formula which allows one to compute the ensemble average of these more general observables?  And second -- and perhaps more importantly -- does the result yield some insights into the structure of the theory and its gravity dual beyond the higher genus partition functions considered in \cite{Maloney:2020nni}?  The answer to the first question is, in fact, not difficult.  The observation begins with the fact (that we will explain in much more detail below) that the counting function for primaries, $\Theta (z_L,z_R,\tau,\bar \tau)$, obeys a version of the heat equation:
 \bea\label{diffusion}
 {\p \Theta \over \p \tau} = {1\over 4\pi i} \nabla_{z_L}^2 \Theta  ~,\qquad  {\p \Theta \over \p \taub} = -{1\over 4\pi i} \nabla_{z_R}^2 \Theta ~.
 \eea
This equation follows from the fact that the stress tensor of a free boson theory is Sugawara, and hence a composite operator quadratic in the $U(1)$ currents; this relates variations with respect to the conformal structure to variations with respect to the $U(1)$ gauge potentials.  By averaging this equation over Narain moduli space we will completely determine the ensemble average of the flavored partition function, a novel (and somewhat less commonly studied) version of the Siegel-Weil formula.

Equation \eqref{diffusion} hints as well towards an answer to our second question, as it allows us to trade conformal structure dependence for dependence on the fugacities.  At the level of the torus partition function this is not particularly interesting, as it simply reflects the fact that the dimensions and spins of primary operators are uniquely given by their $U(1)$ charges
\be
\Delta = Q \cdot Q + \bar Q \cdot  \bar Q \, , \qquad
j = Q \cdot Q - \bar Q \cdot  \bar Q ~.
\ee
The situation at higher genus is considerably more interesting since   the dependence of the higher genus partition function on conformal structure encodes not just the dimensions and spins of primary states but also the operator product expansion coefficients.  In a free CFT, however, the OPE coefficients are completely determined by charge conservation
\be
C_{Q_1,Q_2,Q_3}  \propto \delta(Q_1+Q_2+Q_3)~.
\ee
Thus one might expect that all of the data of a higher genus partition function can be completely packaged into information about the corresponding conserved charges.  Indeed, we will see that this is the case by writing down a higher genus version of the heat equation \eqref{diffusion}.
An interesting feature of this result is that it is possible to go to the boundary of moduli space where a higher genus surface degenerates into a disjoint union of tori.  The result is that all of the data contained in a genus $g$ partition function can be repackaged into the data of the $g^{\rm th}$ moment of the (flavored) torus partition function:
\be
\langle Z(\tau_1,z_1)  Z(\tau_2,z_2) \cdots  Z(\tau_g,z_g)   \rangle \leftrightarrow \vev{Z_g(\tau)}~.
\ee
The averages of these quantities are given by appropriate Eisenstein series, just as in the unflavored case.
In a sense, therefore, this perspective allows us to completely dispense with the higher genus partition functions and consider only statistical properties of the torus partition function.\footnote{This may provide an interesting perspective on the analogy between sphere packing and the modular bootstrap described in \cite{Hartman:2019pcd,Afkhami-Jeddi:2020hde}.  The natural question following \cite{Afkhami-Jeddi:2020hde} is: what is the sphere packing analogue of the conformal bootstrap constraints which go beyond torus modular invariance, such as higher genus modular symmetry or the crossing symmetry of local correlation functions?  Our considerations suggest the following answer: modular properties of higher moments of the theta series appearing in the  sphere packing problem.}
An additional interesting feature of our result is that it allows us to easily compute explicit expressions for the averaged density of states $\langle \rho(\Delta, j, Q^I)\rangle$ and the two point function $\langle  \rho(\Delta_1, j_1, Q_1^I) \rho(\Delta_1, j_1, Q_1^J)\rangle$; it turns out that by including dependence on charge, one finds expressions which are considerably simpler than those which have previously appeared in the literature.

Turning to the holographic interpretation, we show that the statement \cite{Maloney:2020nni,Afkhami-Jeddi:2020ezh} that the averaged partition function can be naturally reproduced in terms of $U(1)^D\times U(1)^D$ Chern-Simons theory generalizes to the flavored case.  The chemical potentials appearing in the flavored partition function map to a choice of boundary conditions in the Chern-Simons theory, in a manner which enforces the proper behavior under modular transformations.

This paper is structured as follows. In Section \ref{sec:flav-Z-intro} we begin with a few remarks on the Narain moduli space and define the averaging procedure for  partition functions.  The flavored partition function on the torus is evaluated in Section \ref{sec:SWformula-torus} using a generalization of the Laplace equation, as well as via a heat equation.  The analysis for the partition function  is generalized to higher genus in Section \ref{sec:higher-g}.   Section \ref{sec:bulk} reproduces the flavored partition function from $U(1)^D \times U(1)^D$ Chern-Simons theory in AdS$_3$.

\section{Flavored partition functions of Narain CFTs}
\label{sec:flav-Z-intro}

In this section we recall aspects of free CFTs in two dimensions, with emphasis on their symmetries and moduli spaces. 

We consider the theory of $D$ real compact bosons $X^I$, $I=1,2, \ldots D$, and its associated $U(1)^D \times U(1)^D$ current algebra. Current algebra primaries are given by the vertex operators\footnote{In string theory language we are setting $\alpha'=2$.}
\be
V_l = e^{il_L \cdot X_L + il_R\cdot X_R}.
\ee
The  momentum vectors $l\equiv (l^I_L, l^I_R)$ live in a lattice $\Gamma$, which has a signature $(D,D)$ inner product:
\be
l \circ l \equiv l_L \cdot l_L - l_R \cdot l_R~.
\ee
The choice of lattice $\Gamma$ labels different possible CFTs, i.e.\,different compactifications of the free bosons.
This choice
is constrained by modular invariance of the torus partition function.
First, invariance under $\tau\to\tau+1$ (i.e.~the quantization of spin) implies that $\Gamma$ is even, i.e. that the vectors $(l^I_L, l^I_R)$ obey 
%
$l \circ l \in 2 \Z$.
Second, invariance under $\tau\to-1/\tau$ implies that $\Gamma$ is self-dual, i.e. that $\Gamma^*=\Gamma$, where the dual lattice $\Gamma^*$ consists of all vectors with integer $\circ$ product with all elements of $\Gamma$.
An even, self-dual lattice of signature $(D,D)$ is known as a Narain lattice.

The eigenvalues of the Virasoro generators $(L_0,\tilde{L}_0)$ are
\bea\label{spectrum0}
L_0 = {1\over 2} l_L^2 +N~,\quad \tilde{L}_0 = {1\over 2}l_R^2 + \tilde{N}~,
\eea
where $N,\tilde{N} \in \Z$ are the integer valued oscillator levels.  
The (unflavored) partition function is
\bea\label{Zgam}
Z_\Gamma(\tau) = {1\over |\eta(\tau)|^{2D} } \sum_{l\in \Gamma} e^{i\pi \tau l_L^2 - i\pi \taub l_R^2}~.
\eea
Here the prefactor counts the oscillator states, i.e. the descendants under the $U(1)^D\times U(1)^D$ current algebra, and the lattice sum counts primaries.
The partition function is modular invariant, in the sense that
\bea
Z_\Gamma(\gamma\tau ) =Z_\Gamma(\tau )~,\quad \gamma =\left({a~b\atop c~d}\right)\in SL(2,\Z)~,\quad \gamma\tau\equiv {a\tau+b\over c\tau+d}~.  
\eea

Given a Narain lattice $\Gamma$, one can always apply an $O(D,D)$ rotation $\Lambda$ to produce another Narain lattice $\Gamma_\Lambda \equiv \Lambda \Gamma$, with $\Lambda \in O(D,D)$.  In fact, it is not hard to show that {\it any} Narain lattice may be obtained by some $O(D,D)$ rotation of a fixed reference lattice $\Gamma_0$.   However, not all such $O(D,D)$ rotations yield distinct CFTs.   First, an $O(D)\times O(D)\in O(D,D)$ rotation will act as a symmetry of a particular theory, since its effect can be undone by a compensating $O(D)\times O(D)$ field redefinition of the fields $(X_L,X_R)$.  The result is that the spectrum of vertex operators and their OPE coefficients will be unchanged by such a rotation.   Second, a subgroup of $O(D,D)$ will leave the lattice $\Gamma$ itself invariant.  This subgroup is just $O(D,D,\Z)$, as can be seen by taking our reference lattice $\Gamma_0$ to be the integer lattice in $\mathbb{R}^{D,D}$.  The result is that the moduli space of inequivalent Narain theories is the coset
\bea\label{modspace}
{\cal M}_D\equiv { O(D,D,\Z)\backslash O(D,D) \slash O(D)\times O(D)}~.
\eea
This moduli space has dimension $D^2$.

This space of free theories can also be described  more explicitly as $\sigma$-models, with action %
\bea
S = {1\over 4\pi} \int\! d^2 \sigma \left( \sqrt{g} g^{\alpha\beta} G_{IJ} \p_\alpha X^I \p_\beta X^J + \eps^{\alpha\beta} B_{IJ} \p_\alpha X^I \p_\beta X^J\right)~.
\eea
Here the boson fields have been scaled to have integer periodicities: $X^I \cong X^I +2 \pi m^I$, $m^I \in \Z$, so the choice of
theory has been packaged into the target space metric $G_{IJ}$ and $B$-field $B_{IJ}$.  These are constant symmetric and antisymmetric matrices, respectively, which can be combined into a $D\times D$ matrix
\bea
E_{IJ} = G_{IJ} +B_{IJ}~.
\eea
One can think of $E$ as a coordinate on the moduli space ${\cal M}_D$.

To understand the Narain moduli space in this language, we introduce the $O(D,D)$ element $g$ that acts on the matrix $E$ as
\bea
g: E \rt gE\equiv (aE+b)(cE+d)^{-1}~,
\eea
where 
\bea
g = \left(\begin{array}{cc} a & b \cr c & d \end{array}\right)~,\quad g^TJg=J~,\quad J = \left(\begin{array}{cc} 0 & I \cr I& 0\end{array} \right)~
\eea
is an element of $O(D,D)$.
Any matrix $E$ is invariant under  some $O(D)\times O(D)$ subgroup of $O(D,D)$.   This can be seen by first noting that $E=I$ is invariant under the action of matrices of the form
\bea
  g = \left(\begin{array}{cc} a & 0 \cr 0 & a \end{array}\right)~,\quad \left(\begin{array}{cc} 0 & b \cr b & 0 \end{array}\right)~,\quad a^Ta = b^T b = I~.
  \eea
The corresponding statement for general $E$ is obtained by conjugating by the action of $O(D,D)$.

To write the spectrum, we introduce the matrix
\bea
M = \left(\begin{array}{cc} G-BG^{-1}B & BG^{-1} \cr
-G^{-1}B  & G^{-1} \end{array}\right)~.
\eea
This is convenient because the $O(D,D)$ rotations act equivariantly on $M$, in the sense that 
\bea\label{buscher}
g: M \rt  gMg^T~.
\eea
Since in the $\sigma$-model formulation the fields have integer periodicities,
the primary states of the theory can be labelled by a vector of integers $(m^I,n_I)$.
In terms of these, the spin $L_0-{\bar L}_0$ and dimension $L_0+{\bar L}_0$ of a given primary state is
\bea
l_L^2-l_R^2 = 2 m^In_I~,\quad l_L^2 + l_R^2 = Z^T MZ~,\quad Z \equiv \left(\begin{array}{c} m^I \cr n_I\end{array} \right)~.
\eea
The T-duality group $O(D,D,\Z)$ is given by those $g$ for which the entries of $g^TZ$ are integer.  In this case the action (\ref{buscher}) is the usual Buscher rule for the T-duality transformation of the target space metric and $B$-field.

It is important to note that the moduli space ${\cal M}_D$ defined above is a homogeneous space which has a unique Riemannian metric which is invariant under an $O(D,D)$ isometry group (generated, in terms of the coset structure, by left multiplication).  This coincides with the usual ``Zamolodchikov'' metric on the CFT moduli space, and is the natural one to use when considering averages over this space of theories.
In particular, we average over moduli space by integrating:
\be
\langle \cdot  \rangle = \frac{1}{Vol({\cal M}_D)} \int_{\cal M_D}  \left(\cdot \right) d\mu
\ee
where $d\mu$ is the associated invariant measure.  We have divided by the volume of ${\cal M}_D$ in order to properly normalize this measure as a probability distribution.  It is important to note that, although $O(D,D)$ has infinite volume, the moduli space ${\cal M}_D$ has finite volume when $D>1$.  This is due to the fact that we have quotiented by the action of the T-duality group; without such a quotient, an interpretation of $d\mu$ as a normalizable probability measure would be impossible.

We wish to study the flavored partition function, which is obtained by introducing a set of 2D chemical potentials $z\equiv (z_L^I,z_R^I)$ that couple to the $U(1)^D\times U(1)^D$ charges of a state.  These charges are just the individual components of  the lattice vector $l=(l^I_L, l^I_R)$, so the flavored partition function is
\bea
Z_\Gamma(\tau,z) =  {1\over |\eta(\tau)|^{2D} } \sum_{l\in \Gamma} e^{i\pi \tau l_L^2 - i\pi \taub l_R^2+2\pi i z_L \cdot l_L -2\pi i z_R \cdot l_R}~.
\eea
We note that only the lattice sum has been modified; the prefactor remains the same, because the action of the $U(1)$ current algebra will not change the charge of a state.

There is one important distinction, which is that the potentials $z$ are not invariant under the $O(D)\times O(D)$ rotations described above.  The reason is easy to understand.
Given a point in moduli space corresponding to a choice of $E$ there is an equivalence class of lattices related by $O(D)\times O(D)$ rotations, all corresponding to the same CFT.   However, in a given CFT there are many possible choices of basis for the $U(1)^D\times U(1)^D$ symmetry algebra, which are related precisely by these $O(D)\times O(D)$ transformations.  When we introduce potentials $z$ we have implicitly made a choice of basis.  So the flavored partition function should be viewed as a function on the space of Narain lattices
$O(D,D,\Z)\backslash O(D,D)$ rather than on the moduli space of CFTs (\ref{modspace}).
This will be important when we consider the average of flavored quantities, because we must now integrate over this larger moduli space.
In particular, we will consider averages of the form
%
%
%
\bea
\langle \cdot \rangle = \frac{1}{{\rm Vol}(O(D,D,\Z)\backslash O(D,D))} \int_{O(D,D,\Z)\backslash O(D,D)} \left(\cdot\right) d\mu
\eea
For quantities which are $O(D)\times O(D)$ invariant (such as unflavored partition functions) this reduces to the average over ${\cal M}_D$ described above.
But this procedure can now be applied to flavored quantities as well.
%

\section[Siegel-Weil formula for flavored partition functions: torus case]{Siegel-Weil formula for flavored partition\\ functions: torus case}
\label{sec:SWformula-torus}

In this section we will compute the average of flavored CFT partition functions on the torus. 
We will do so by showing that it satisfies a set of differential equations, combined with knowledge of its behavior at the boundary of moduli space.
We will begin with a review of the unflavored case, before describing two differential equations -- both a 	``Laplace equation'' and a ``heat equation'' -- obeyed in the flavored case.  This latter equation in particular will allow us to easily reduce the computation of the averaged flavored partition function to the unflavored case.

\subsection{The flavorless Siegel-Weil formula}

We begin by describing the Laplace equation obeyed by the partition function, which was used by \cite{Maloney:2020nni} to derive the Siegel-Weil formula in the unflavored case.  We will present a streamlined derivation of this equation in a form which can be easily adapted to the flavored case.

We start by writing the partition function as
\bea
Z_\Gamma(\tau) = {1\over |\eta(\tau)|^{2D} } \Theta_\Gamma(\tau),~~~~~\Theta_\Gamma(\tau)\equiv \sum_{l\in \Gamma}  Q(l,\tau)
\eea
where  
\bea
 Q(l,\tau)\equiv e^{i\pi \tau l_L^2 - i\pi \taub l_R^2}=e^{i\pi \tau_1( l_L^2 - l_R^2)} e^{-\pi\tau_2 (l_L^2+l_R^2) }~.
\eea
with $\tau=\tau_1+i\tau_2$, $\taub=\tau_1-i\tau_2$.
We have separated out the theta function $\Theta_\Gamma(\tau)$ which counts primary states.
We denote the Laplacian acting on the modular parameter $\tau$ as
\bea
\Delta_{\cal H}   = -\tau_2^2 \left( {\p^2 \over \p \tau_1^2 }+{\p^2 \over \p \tau_2^2}\right) = -4\tau_2^2 {\p^2 \over \p \tau \p \taub }~.
\eea
It is then straightforward to check that
\bea
\Delta_{\cal H}Q(l,\tau) = -4\pi^2 \tau_2^2 l_L^2 l_R^2Q(l,\tau)~.
\eea

We now consider the Laplacian $\Delta_{\cal M}$ acting on the moduli space of Narain lattices.  While we could write this operator  in terms of the $(G_{IJ},B_{IJ})$ target space fields, it is simpler to think of this Laplacian as an operator on the $O(D,D)$ group manifold.
Since $Q(l,\tau)$ is invariant under $O(D)\times O(D)$ rotations, these two versions of the Laplacian will be proportional to one another.
We start by defining $O(D,D)$ as the linear transformations which preserve the quadratic form $\eta_{AB}Z^A Z^B$ where $A,B = 1,2,\ldots , 2D$ and  $\eta_{AB} = {\rm diag}(1^D, -1^D)$.
Writing the $O(D,D)$ generators as
\bea
J^{AB} =    \eta^{BC} Z^A {\p \over \p Z^C}-  \eta^{AC} Z^B {\p \over \p Z^C}~
\eea
the quadratic Casimir is
\bea
J^2 = \eta_{AC}\eta_{BD} J^{AB}J^{CD}~.
\eea
We now use the fact that the charge vector $l=(l_L^I,l_R^I)$ transforms as a (contravariant) vector under the $O(D,D)$ rotations.
In particular, we can assemble these charges into an $O(D,D)$ vector $Z^A$ as:
\bea
l_L^I= Z^I~,\quad l_R^I = Z^{D+I}~,\quad I=1,2, \ldots D~.
\eea
%
%
Since the quadratic Casimir is proportional to the Laplacian, this provides an explicit expression for the Laplacian as a differential operator.

Explicitly, when acting on functions of the charge vector the quadratic Casimir takes the form
\bea
J^2 = L^I_J L^I_J + R^I_J R^I_J  +2 T^I_J T^I_J~,
\eea
with
\bea\label{LRdef}
  L^I_J = l_L^I {\p \over \p l_L^J}-  l_L^J {\p \over \p l_L^I} ~,\quad  R^I_J = l_R^I {\p \over \p l_R^J}-  l_R^J {\p \over \p l_R^I}~,\quad T^I_J = l_L^I {\p \over \p l_R^J} +  l_R^J {\p \over \p l_L^I}~.
\eea
%
%
Since the $l_{L,R}^2$ are annihilated by $L^I_J$ and $R^I_J$, and $l_L^2-l_R^2$ is annihilated by all the generators,  we have
\bea
J^2 Q(l,\tau) = e^{i\pi \tau_1( l_L^2 - l_R^2)}   \times  \left[ 2 T^I_J T^I_J  e^{-\pi\tau_2 (l_L^2+l_R^2) } \right]~.
\eea
An elementary computation   yields
\begin{align}\label{Tcomp}
 e^{i\pi \tau_1( l_L^2 - l_R^2)} 2T^I_J T^I_J  e^{-\pi\tau_2 (l_L^2+l_R^2) } &= 8\left[ 4\tau_2^2 {\p^2 \over \p \tau \p \taub}+D\tau_2 {\p \over \p \tau_2}\right] e^{i\pi \tau l_L^2 -i\pi \taub l_R^2 }  \cr
 & = -8 \left[ \Delta_{\cal H} -D\tau_2 {\p \over \p \tau_2}\right]  e^{i\pi \tau l_L^2 -i\pi \taub l_R^2 }  ~.
\end{align}
%
%
%
We will normalize our Laplacian as 
\bea
\Delta_{\cal M} =- {1\over 8}J^2
\eea
to match \cite{Maloney:2020nni}, so that our result reads
\bea\label{Qdfiff}
 \left[ \Delta_{\cal H} -D\tau_2 {\p \over \p \tau_2}-\Delta_{\cal M} \right]  Q(l,\tau)=0~.
\eea

We will rewrite this as
\bea\label{ldif}
\left[ \Delta_{\cal H} +s(s-1)-\Delta_{\cal M} \right] \left(\tau_2^{D/2} Q(l,\tau)\right) =0~,\quad s\equiv D/2~.
\eea
We can now sum this over $\Gamma$ to conclude that the theta function obeys the same differential equation:
\be\label{prim}
\left[ \Delta_{\cal H} +s(s-1)-\Delta_{\cal M} \right] \left(\tau_2^{D/2} \Theta_\Gamma(\tau)\right) =0~.
\ee
In this expression $\Delta_{\cal M}$ is now the Laplacian on the space of Narain lattices $\Gamma$. 
We note that, since $|\eta(\tau)|^{-2D}$ and $\tau_2^{D/2}$ have the same modular transformation properties, $\tau_2^{D/2} \Theta_\Gamma(\tau)$ is modular invariant

We now integrate this equation over the moduli space ${\cal M}_D$ to obtain an equation for the object
\be
H(\tau)\equiv \tau_2^{D/2} \langle \Theta_\Gamma(\tau)\rangle~.
\ee
The crucial observation is that, since $\Delta_{\cal M} \Theta_\Gamma(\tau)$ is a total derivative on ${\cal M}_D$, its  integral vanishes.\footnote{To argue that this one must in addition show that the surface terms arising on the boundary of ${\cal M}_D$ vanish.  It is easy to see that this occurs when $D>2$ by considering the explicit behavior of the lattice sum; see \cite{Maloney:2020nni} for details.}
The result is that $H(\tau)$ is a modular invariant eigenfunction of the Laplacian on the upper half plane:
\bea
\left[ \Delta_{\cal H} + s(s-1) \right] H(\tau)=0~.
\eea
One solution to this equation is the Eisenstein series\footnote{To see that this is an eigenfunction of the Laplacian we note that $\tau_2^{D/2}$ is itself an eigenfunction of $\Delta_{\cal H}$ with the correct eigenvalue.  The Eisenstein series is the sum of this eigenfunction over the modular group, $E(\tau)= \sum_{\gamma \in SL(2,\Z)/\Z} \gamma\tau_2^{D/2}$, which gives a modular invariant eigenfunction with the same eigenvalue.  Here the subgroup $\Z$ is the set of matrices $\left({1~n\atop0~1}\right)$ which leave $\tau_2$ invariant.  The coset $SL(2,\Z)/\Z$ can then labelled by pairs of coprime integers $(c,d)$ which make up the lower row of an $SL(2,\Z)$ matrix, giving the form of the Eisenstein series given in equation (\ref{eis}).}
\bea\label{eis}
E_{D/2}(\tau) \equiv \tau_2^{D/2} \sum_{(c,d)=1} {1\over |c\tau+d|^D}~.
\eea
We can now argue that in fact $H=E(\tau)$.  One way to do so is to note that, since we are considering modular invariant functions, we can effectively view this as a Laplace equation on the fundamental domain ${\cal H}/SL(2,\Z)$ which has finite volume.  Compactifying the fundamental domain by adding the point at infinity ($\tau=i\infty$), we can use the uniqueness of solutions to the Laplace equation with negative eigenvalue.  One only has to check that $H(\tau)$ and $E_{D/2}(\tau)$ have the same behavior as $\tau\to\infty$.  Putting this together gives the Siegel-Weil formula for the torus partition function:
\be\label{unFlavored-Z}
\langle Z_\Gamma (\tau) \rangle = \frac{\tau_2^{D/2}}{|\eta|^{2D}} \sum_{(c,d)=1} {1\over |c\tau+d|^D}~.
\ee

\subsection{Flavored Laplace equation}

We now extend this to the flavored partition function, which we write as
\bea
Z_\Gamma(\tau,z) =  {1\over |\eta(\tau)|^{2D} } \Theta_\Gamma(\tau,z),\quad \Theta_\Gamma(\tau,z) \equiv \sum_{l\in \Gamma}P(l,z,\tau),
\eea
where again the function $\Theta_\Gamma(\tau,z)$ counts the contribution of primary states, and
\bea
P(l,z,\tau)=  e^{i\pi \tau l_L^2 - i\pi \taub l_R^2+2\pi i z_L \cdot l_L -2\pi i z_R \cdot l_R}~.
\eea
The flavored partition function is not modular invariant, but instead transforms covariantly as
\begin{align}\label{Zmod}
Z_\Gamma\left(\frac{a\tau+b}{c\tau+d},\frac{a\btau+b}{c\btau+d},\frac{z_L^I}{c\tau+d},\frac{z_R^I}{c\btau+d}\right)
=\exp \left[ \frac{ic\pi  z_L^2}{c\tau+d} - \frac{ic\pi  z_R^2}{c\btau+d}  \right]Z_\Gamma(\tau,\btau,z_L^I,z_R^I)~.
\end{align}

We will begin by deriving a version of the Laplace equation.
In our derivation of the Laplace equation above, we used the fact that the charge vector $l=(l_L^I, l_R^I)$ could be packaged into a contravariant vector under $O(D,D)$ transformations.
Similarly, the chemical potentials $z=(z_L^I,z_R^I)$ can be assembled into a covariant vector under $O(D,D)$, since the inner product $ z_L \cdot l_L - z_R \cdot l_R$ is $O(D,D)$ invariant.  The result is that the $O(D,D)$ generators (\ref{LRdef}) will, when acting on functions of both the charge vector $l$ and the chemical potential $z$, take the form
\begin{align}\label{lrt}
L^I_J & =  l_L^I {\p \over \p l_L^J}-  l_L^J {\p \over \p l_L^I} - z_L^I {\p \over \p z_L^J}+  z_L^J {\p \over \p z_L^I} \cr
 R^I_J &= l_R^I {\p \over \p l_R^J}-  l_R^J {\p \over \p l_R^I} -z_R^I {\p \over \p z_R^J}+  z_R^J {\p \over \p z_R^I} \cr
 T^I_J &= l_L^I {\p \over \p l_R^J} +  l_R^I {\p \over \p l_L^J}+ z_L^J {\p \over \p z_R^I} +  z_R^J {\p \over \p z_L^I}~.
\end{align}
The quadratic Casimir on $O(D,D)$ will again take the form
\bea
\hat{J}^2 =  L^I_J L^I_J + R^I_J R^I_J  +2 T^I_J T^I_J~,
\eea
where the hat indicates that this quadratic Casimir is now understood as a differential operator on functions of both $l$ and $z$.

We now follow the same logic as in our derivation of equation (\ref{ldif}).   Of course, the generators (\ref{lrt}) all annihilate the $O(D,D)$ invariant combination $z_L \cdot l_L - z_R \cdot l_R$ which appears in our expression for $P(l,z,\tau)$.  The result is that the computation  reduces to the one described earlier, and we find
\bea\label{fldif}
\left[ \Delta_{\cal H} +s(s-1)-\Delta_{\cal M}\right] \left(\tau_2^{D/2} P(l,z,\tau)\right) =0~,\quad s= D/2~,
\eea
where again $\Delta_{\cal M} =-{1\over 8} \hat{J}^2 $.
As before, we can sum over the lattice $\Gamma$ to see that:
\be\label{lap2}
\left[ \Delta_{\cal H} +s(s-1)-\Delta_{\cal M} \right] \left(\tau_2^{D/2} \Theta_\Gamma(\tau,z)\right)=0 . 
\ee
We note that the Laplacian $\Delta_{\cal M}$ appearing here is an $O(D,D)$ Laplacian which now acts on both the space of Narain lattices $\Gamma$ as well on the vector $z=(z_L^I,z_R^I)$ of chemical potentials.

We now wish to integrate this equation over the space of Narain lattices $\Gamma$ to obtain an equation for
\bea
G(z,\tau) \equiv  \tau_2^{D/2} \langle \Theta_\Gamma(\tau,l,z)\rangle~.
\eea
It is important to remember that -- as described in section 2 -- the average $\langle \cdot \rangle$ should be understood as an integral over all of $O(D,D)$ rather than just $O(D,D)/O(D)\times O(D)$.
This implies that the integrated expression $G(z,\tau)$ will only depend on the potentials through the $O(D)\times O(D)$ invariant combinations $z_L^2$ and $z_R^2$.
Thus in evaluating this integral many of the terms appearing in $\Delta_{\cal M} =-{1\over 8} \hat{J}^2$ will vanish.  In particular,
$z_L^2$ and $z_R^2$ are annihilated by the $z$-dependent terms in $L^I_J$ and $R^I_J$.    Therefore, all that survives from $\hat{J}^2$  is the purely  $z$-dependent contribution from $2T^I_JT^I_J$.\footnote{We have also used the fact that the terms in $\hat{J}^2$ involving only $l_{L,R}$ will lead to boundary terms which vanish upon integrating over $O(D,D)$ when $D>2$, exactly as in the unflavored case.}  The result is that 
\bea\label{avgfdif0}
\left[ \Delta_{\cal H} +s(s-1)+{1\over 4} \left( z_L^J {\p \over \p z_R^I} +  z_R^J {\p \over \p z_L^I} \right)  \left( z_L^J {\p \over \p z_R^I} +  z_R^J {\p \over \p z_L^I} \right)  \right]G(z,\tau) =0~.
\eea
This is the version of the Laplace equation which is obeyed by the average of the flavored partition function.

One obvious solution to this equation can be obtained by generalizing the Eisenstein series (\ref{eis}) in order to accommodate the more general modular transformation rule (\ref{Zmod}):
%
\bea\label{Gsol}
G(z,\tau) = \tau_2^{D/2} \sum_{(c,d)=1} {e^{ -i \pi \left( {cz_L^2 \over c\tau+d}-{cz_R^2 \over c\tau+d} \right) }\over |c\tau+d|^{D}}~.
\eea
This is a solution to the equation with the correct transformation properties \eqref{Zmod}.  We will argue below that this is indeed the correct answer for the average of the flavored partition function.  However, we note that equation (\ref{avgfdif0}) has many other solutions as well; for example, we can multiply it by any function of $z_L^2-z_R^2$.   We will therefore give an alternative argument based on the  ``heat equation'' obeyed by the flavored partition function.

\subsection{Heat equation}

The starting observation for our derivation is that equation (\ref{spectrum0}) implies that the conformal weights $L_0$ and ${\bar L}_0$ are determined by the charge vector $l=(l_L^I, l_R^I)$.  This implies that the $(\tau, {\bar \tau})$ dependence of the flavored partition function is be determined its  $z=(z_L^I, z_R^I)$ dependence.
In particular, we note that 
\bea
P(l,z,\tau)\equiv  e^{i\pi \tau l_L^2 - i\pi \taub l_R^2+2\pi i z_L \cdot l_L -2\pi i z_R \cdot l_R}~,
\eea
obeys
\bea
 {\p \over \p \tau} P(l,z,\tau) = {1\over 4\pi i} \nabla_L^2 P(l,z,\tau) ~,\quad  {\p \over \p \taub} P(l,z,\tau) = -{1\over 4\pi i} \nabla_R^2 P(l,z,\tau)~,
 \eea
with
\bea
\nabla_L^2 =  {\p^2 \over \p z_L^I \p z_L^I}~,\quad  \nabla_R^2 =  {\p^2 \over \p z_R^I \p z_R^I}~.
\eea
Summing over lattice points, we conclude that $\Theta_\Gamma(\tau,z)$ will obey the same equation.

This heat equation is obeyed by the partition function for every CFT in the Narain ensemble.
We can therefore integrate over Narain lattices, and conclude that
\bea
Y(z_L,z_R,\tau,\taub)\equiv \langle \Theta_\Gamma(\tau,z)\rangle~
\eea
obeys the same equation.
We note that, as in the previous section, the integration over $O(D,D)$ implies that the only dependence on potentials is through the $O(D)\times O(D)$ invariants $z_L=\sqrt{z_L^I z_L^I}$ and $z_R=\sqrt{z_R^I z_R^I}$.  Writing the Laplace operators $\nabla_{L,R}^2$ in spherical  coordinates and discarding the angular parts, the heat equation becomes
\bea\label{heateq2}
{\p Y \over \p \tau} = {1\over 4\pi i} \left[ {\p^2 Y \over \p z_L^2} +{D-1\over z_L}{\p Y \over \p z_L}\right]~,\quad {\p Y \over \p \taub} = -{1\over 4\pi i} \left[ {\p^2 Y \over \p z_R^2} +{D-1\over z_R}{\p Y \over \p z_R}\right]~.
\eea

These heat equations can be used to fix $Y(z_L,z_R,\tau,\taub)$.   First, from the unflavored analysis we know that
\bea
Y(z_L=0,z_R=0,\tau,\taub)  = \sum_{(c,d)=1} {1\over |c\tau+d|^{D}}~.
\eea
From its definition, we know that it is possible to make the power series expansion $Y(z,\tau)  = \sum_{m,n=0}^\infty Y_{m,n}(\tau,\taub) z_L^{2m} z_R^{2n}$.   
The heat equation then provides a set of recursive relations among the coefficients in this series expansion.
This is easiest to implement by writing
\bea
Y(z_L,z_R,\tau,\taub) =  \sum_{(c,d)=1} {e^{ -i \pi \left( {cz_L^2 \over c\tau+d}-{cz_R^2 \over c\tau+d} \right) }\over |c\tau+d|^{D}}+\Delta Y(z_L,z_R,\tau,\taub) ~.
\eea
The first term on the right hand side obeys the heat equations, so
$\Delta Y$ must as well. It will therefore obey the heat equations and admit the expansion $Y(z ,\tau)  = \sum_{m,n=0}^\infty \Delta Y_{m,n}(\tau,\taub) z_L^{2m} z_R^{2n}$ with $\Delta Y_{0,0}=0$.  It is simple to see that the heat equation implies recursion relations that force all of $\Delta Y_{m,n}=0$.   We conclude
\bea\label{ZFavg}
Y(z_L,z_R,\tau,\taub) =  \sum_{(c,d)=1} {e^{ -i \pi \left( {cz_L^2 \over c\tau+d}-{cz_R^2 \over c\tau+d} \right) }\over |c\tau+d|^{D}}~.
\eea
We saw in the previous section that this also obeys the flavored Laplace equation, as it should. The above object is often called the non-holomorphic Jacobi-Eisenstein series; to our knowledge, holomorphic versions of this quantity were first considered in \cite{arakawa1994jacobi}.

\subsection{Average density of states}

Our conclusion is that the average flavored partition function is
\be
\langle Z(\tau,z)\rangle 
= {1\over |\eta(\tau)|^{2D} } \sum_{(c,d)=1} {e^{ -i \pi \left( {cz_L^2 \over c\tau+d}-{cz_R^2 \over c{\bar \tau}+d} \right) }\over |c\tau+d|^{D}}~.
\ee
We wish to extract from this formula the averaged density of states $ \rho(j,\Delta, Q^I,{\bar Q}^I)$, as a function of the spin
$j=(L_0-{\bar L}_0)\in \Z$, dimension $\Delta=L_0+{\bar L}_0$ and charges $Q^I, {\bar Q}^I$ of a state.
As our partition function depends only on the $O(D)\times O(D)$ invariants $z_L^2$ and $z_R^2$, the resulting density of states is a function only of the total charges $Q = \sqrt{Q^I Q^I}$ and ${\bar Q}= \sqrt{{\bar Q}^I  {\bar Q}^I}$ in the left- and right-moving sectors, respectively.
As we are interested only in the density of primary states, we will omit the prefactor $|\eta(\tau)|^{-2D}$ in what follows.

We begin by
noting that the term in the sum with $(c,d)=(0,1)$ simply describes the contribution of the ground state.  We will therefore concentrate on the terms in the sum with $c>0$.
To extract the average density of states, we will first perform the Fourier transform which takes us from a sector of fixed chemical potentials $(z^I_L,z^I_R)$ to a sector of fixed charges $(Q^I,{\bar Q}^I)$, where the contribution to the partition function is:
\be
Z(\tau, Q^I, {\bar Q}^I) = \int dz_L^I dz_R^I~
e^{2\pi i (z_L^I Q^I +z_R^I {\bar Q}^I)}
\left(\sum_{(c,d)=1} {e^{ -i \pi \left( {cz_L^2 \over c\tau+d}-{cz_R^2 \over c\overline{\tau}+d} \right) }\over |c\tau+d|^{D}}\right)~.
\ee
The usual unflavored torus partition function can be obtained by integrating this expression with measure $dQ^I d{\bar Q}^I$.
Now, the integrals over $(z_L^I,z_R^I)$ are straightforward $D$-dimensional Gaussian integrals.  These cancel out the factor of $|c\tau+d|^D$ in the denominator to give:
\bea
Z(\tau, Q^I,{\bar Q}^I) = e^{-2\pi \tau_2\left(Q^2+{\bar Q}^2\right)} \sum_{(c,d)=1}c^{-D} e^{2\pi i \left(Q^2-{\bar Q}^2\right) (\tau_1+d/c)}~.
\eea
We now let $d=d^* + n c$ and replace the sum over $d$ with a sum over $n\in \Z$ and a sum over the integers $0\le d^*<c$ which are coprime to $c$.
The sum over $n$ is:
\be
\sum_{n\in\Z} e^{2\pi i n(Q^2-{\bar Q}^2)} = \sum_{j\in \Z} \delta\left(j-(Q^2-{\bar Q}^2)\right)~,
\ee
which gives
\bea \label{z-r}
Z(\tau, Q^I,{\bar Q}^I) = e^{-2\pi \tau_2\left(Q^2+{\bar Q}^2\right)} \sum_{j\in\Z} \delta\left(j-(Q^2-{\bar Q}^2)\right) e^{2\pi i j \tau_1}\sum_{c=1}^\infty c^{-D} \left(\sum_{d^*} e^{2\pi i j d^*/c}\right)~.
\eea
We recognize $\Delta=Q^2+{\bar Q}^2$ and $j=Q^2-{\bar Q}^2$ as the dimension and spin of a primary state, as expected.
The quantity in the parenthesis is known as Ramanujan's sum, and is usually denoted:
\be
c_c(j) \equiv \sum_{d^*}e^{-2\pi i j d^*/c}~.
\ee
The sum over $c$ in the expression \eqref{z-r} can be computed, and the result is a factor we will call:
\be
\kappa(j,D)\equiv\sum_{c=1}^\infty \frac{c_c(j)}{c^{D}} = \left\{
{\frac{\sigma_{D-1}(j)}{j^{D-1} \zeta(D)}~,~~{\rm if}~j\ne 0
\atop
\frac{\zeta(D-1)}{\zeta(D)}~,~~~~{\rm if}~{j=0}}\right.
\ee
We can now take the inverse Laplace transform in the $\tau_2$ variable to extract the density of states:
\be\label{density}
\rho(j,\Delta, Q^I, {\bar Q}^I) =  \kappa(j,D)\,\, \delta\left(\Delta-(Q^2+{\bar Q}^2)\right)\, \delta\left(j-(Q^2-{\bar Q}^2)\right)~.  
\ee
This is our final formula for the averaged density of states.  As anticipated, it depends only on the total charges $Q$ and ${\bar Q}$, and these total charges are related to dimension and spin in the usual way.

To understand this formula, it is useful to compare this to the expression the total density of states in the averaged Narain theory.  To do so, we simply integrate this over the space of charges $(Q^I,{\bar Q}^I)$ to give the total density of states:
\bea
\rho(j,\Delta) &=& \kappa(j,\Delta) \int dQ^I d{\bar Q}^I  \delta\left(\Delta-(Q^2+{\bar Q}^2)\right)\delta\left(j-(Q^2-{\bar Q}^2)\right)
\cr
&=& \kappa (j,\Delta) \left(\frac{2 \pi^D}{\Gamma\left(\frac{D}{2}\right)^2}\right) (\Delta^2-j^2)^{D/2-1}~.
\eea
The second line involves a Jacobian factor as well as the volumes of $D$-dimensional spheres in charge space of radius $Q=\sqrt{(\Delta+j)/2}$ and ${\bar Q}=\sqrt{(\Delta-j)/2}$, respectively.  This expression matches precisely the averaged density of states in the Narain theory derived in \cite{Afkhami-Jeddi:2020ezh}.

In retrospect, we could have derived our expressions for the average flavored partition function using a somewhat different logic.  In particular, we could have started with the observation that the average density of states $\rho(j,\Delta, Q^I, {\bar Q}^I)$ must be a function only of $Q=\sqrt{Q^I Q^I}$ and ${\bar Q}=\sqrt{{\bar Q}^I {\bar Q}^I}$, and that these are completely determined by that dimension and spin using the usual formulas
$\Delta=Q^2+{\bar Q}^2$ and $j=Q^2-{\bar Q}^2$.  Equation (\ref{density}) is then the only possible form of the density of states which is consistent with the known expression for the total density of states appearing in \cite{Afkhami-Jeddi:2020ezh}.\footnote{Indeed, our derivation based on the heat equation in the previous subsection -- as it is similarly a simple consequence of $\Delta=Q^2+{\bar Q}^2$ and $j=Q^2-{\bar Q}^2$ --could be considered a different version of this argument.}


\subsection{The $\tau\to 0$ limit}\label{tau0}

We have emphasized the use of the heat equation in fixing the form of flavored partition functions.   In familiar physical systems in which a heat equation arises one is usually interested in solutions with specified boundary conditions at some initial time. In our context $\tau$ plays the role of time.  As a choice of initial time we here consider the case $\tau \rightarrow 0$, at which the flavored partition function takes a distributional form which can be computed fairly explicitly.  Our point in this section  simply is to note that the partition function at generic $\tau$ can be recovered from this singular limit by using the heat equation.

We focus on the factor in the partition function counting primaries,
\begin{align}\label{primary-Z}
Y_D=\sum_{(c,d)=1} \frac{e^{-\pi i   \left( \frac{cz^2}{c\tau+d}-\frac{c\bar{z}^2}{c\Bar{\tau}+d}\right)}}{ |c\tau+d|^D}~.
\end{align}
We consider the $d=0$ term first. For odd $D$, we can express this term as a derivative of the Dirac delta function as follows
\begin{align}
Y_{D,\text{odd}}^{(d=0)} &=  \frac{e^{-\pi i   \left( \frac{z^2}{\tau}-\frac{\Bar{z}^2}{\Bar{\tau}}\right)}}{ |\tau|^D} 
= \frac{1}{\pi^{D-1}} (\pd_{z^2}\pd_{\bar z^2})^{D-1\over 2}\frac{e^{-\pi i   \left( \frac{z^2}{\tau}-\frac{\Bar{z}^2}{\Bar{\tau}}\right)}}{ |\tau|} 
\cr & \stackrel{\tau\to 0}{\to}   \frac{1}{\pi^{D-1}}~\left[(\pd_{z^2})^{D-1\over 2} \delta(z)\right]\left[( \pd_{\bar z^2})^{D-1\over 2}\delta(\bar z) \right]~.
\end{align}
On the other hand, for even $D$ this is
\begin{align}
Y_{D,\text{even}}^{(d=0)} &=  \frac{e^{-\pi i   \left( \frac{z^2}{\tau}-\frac{\Bar{z}^2}{\Bar{\tau}}\right)}}{ |\tau|^D} 
= \frac{1}{\pi^{D-2}|\tau|} (\pd_{z^2}\pd_{\bar z^2})^{D-2\over 2}\frac{e^{-\pi i   \left( \frac{z^2}{\tau}-\frac{\Bar{z}^2}{\Bar{\tau}}\right)}}{ |\tau|} 
\cr &
\stackrel{\tau\to 0}{\to}   \frac{1}{\pi^{D-2}|\tau|}~\left[(\pd_{z^2})^{D-2\over 2} \delta(z)\right]\left[( \pd_{\bar z^2})^{D-2\over 2}\delta(\bar z) \right]~.
\end{align}
For the terms with $d\neq 0$, we have the following sum over co-primes
\begin{align}
Y_D^{(d\neq 0)}=\ \sum_{(c,d)=1}\hspace{-.2cm}{}^{'}\, \frac{e^{-\pi i   \left( \frac{c z^2}{c\tau+d}-\frac{c\Bar{z}^2}{c\Bar{\tau}+d}\right)}}{ |c\tau+d|^D}~.
\end{align}
Here the prime indicates that we consider terms with $d\neq 0$.
Next, we can write $c$ as $j+ kd$, with $j=0,\cdots, d-1$ with the co-prime condition $(j,d)=1$. Setting $\tau=0$, we get
\begin{align}
Y_D^{(d\neq 0)}\big|_{\tau\to 0} =  \sum_{d=1}^{\infty}  \sum_{k=-\infty}^{\infty} \exp\left[-\pi i k (z^2 -\bar z^2)\right]\sum_{\substack{j=0\\(j,d)=1}}^{d-1}\exp\left[-\pi i \frac{j}{d} (z^2 -\bar{z}^2)\right] d^{-D}~.
\end{align}
The sum over $k$ above gives the Dirac comb, $\Sha(x)=\sum_{a=-\infty}^\infty \delta(x-a)$, while the
 sum over $j$ is the Ramanujan sum
\begin{align}
Y_D^{(d\neq 0)}\big|_{\tau\to 0} =   ~\Sha  (\tfrac{\bar z^2- z^2}{2}) ~ \sum_{d=1}^{\infty}
c_{d}(\tfrac{\bar z^2- z^2}{2})
~ d^{-D}~.
\end{align}
Performing the sum over $d$ yields the result
\begin{align}
Y_D^{(d\neq 0)}\big|_{\tau\to 0} =
\frac{\sigma_{D-1}(\tfrac{\bar z^2-z^2}{2})}{(\tfrac{\bar z^2-z^2}{2})^{D-1} \zeta(D)} \Sha   (\tfrac{\bar z^2-z^2}{2})~.
\end{align}
where, $\sigma_r(s)$ is the divisor function and $\zeta(p)$ is the Riemann-zeta function.
Hence, the primary counting partition function at $\tau\to0$ is
\begin{align}\label{comb}
 ~Y_D|_{\tau=0}~=~\frac{1}{\pi^{D-2+m} |\tau|^{1-m}}
 \left|(\pd_{z^2})^{D-2+m\over 2} \delta(z)\right|^2
 +
 \frac{\sigma_{D-1}(\tfrac{\bar z^2-z^2}{2})}{(\tfrac{\bar z^2-z^2}{2})^{D-1} \zeta(D)} \Sha   (\tfrac{\bar z^2-z^2}{2})~.
\end{align}
where, $m=(D\!\!\mod 2)$.

A natural question is: why does the contribution to the partition function localize to the points where $\bar z ^2 -z^2$ is an even integer? This can be understood by considering the unaveraged $D=1$ theory, where a similar phenomenon happens.
The primary counting partition function at $\tau\to0$ is
\begin{align}
Y_1(R)\big|_{\tau\to 0} = {\rm Tr}[y^{J_0} \bar y^{\bar J_0}]&=\sum_{n,w} e^{2\pi i {z} \left(\frac{n}{R}+\frac{w R}{2}\right)} e^{-2\pi i \bar{z} \left(\frac{n}{R}-\frac{w R}{2}\right)} \nn \\
&= \sum_{n} e^{2\pi i ({z} -\bar{z})\frac{n}{R} }\sum_{w}e^{2\pi i{({z} +\bar{z})} \frac{R w}{2}}
= \Sha\left(\frac{{z}-\bar {z}}{R}\right)\Sha\left(\frac{({{z}+\bar {z}}){R}}{2}\right)
\end{align}
We see that the partition function localises to the points where
\begin{align}
\frac{{z}-\bar {z}}{R} \in \mathbb{Z}, \ \frac{({{z}+\bar {z}}){R}}{2} \in \mathbb{Z}
\end{align}
Combining the above two conditions gives the weaker condition
\begin{align}
\frac{{z}^2 -\bar{z}^2}{2} \in \mathbb{Z}
\end{align}
which is independent of $R$ and is the same condition enforced by the Dirac comb appearing in \eqref{comb}.

\section{Higher genus}
\label{sec:higher-g}

The results of the previous section can be generalized in a reasonably straightforward way to the partition function on higher genus surfaces.
In particular, we will show that a flavored version of the genus $g$ partition function obeys analogs of the Laplace and heat equations described above.
This will lead to a similar formula for the average flavored partition function.

The partition function of a Narain CFT on a Riemann surface $\Sigma$ of genus $g$ is
\bea\label{Zg}
  Z_{g,\Gamma}(\tau)= \frac{1}{\Phi(\tau)} \theta_\Gamma(\tau)~,
  \eea
where the Siegel-Narain theta function is
\bea\label{thetag2}
\theta_\Gamma(\tau) =  \sum_{l^i\in \Gamma} e^{i\pi \tau_{ij} l_L^i\cdot l_L^j - i\pi \taub_{ij} l_R^i\cdot l_R^j}~.
\eea
The prefactor $\Phi(\tau)$ comes from the integral over oscillator modes, and can be expressed in terms of the one-loop determinant of the scalar Laplacian on $\Sigma$.  This contribution is independent of the Narain lattice $\Gamma$, so will not be important in what follows.\footnote{$\Phi(\tau)$ does, however, depend on the central charge and is necessary in order to obtain the correct behavior under Weyl transformations.}
The period matrix  $\tau_{ij}$ is a complex, symmetric $g\times g$ matrix with positive imaginary part; i.e. $\tau_{ij}$ lives in the Siegel upper half-space ${\cal H}_g$.   Not every such matrix is actually the period matrix of some Riemann surface, but (\ref{Zg}) is well defined regardless.
%

The flavored partition function is now obtained by introducing a set of chemical potentials $z=(z_{Li}^{I},z_{Ri}^{I})$ with $i=1\dots g$ and $I=1\dots D$.  These measure the charges that propagate around the various cycles in the Riemann surface,
and can be understood as holonomies for background $U(1)^D$ Wilson lines.
  The flavored partition function is
\be
  Z_{g,\Gamma}(\tau,z)= \frac{1}{\Phi(\tau)} \theta_\Gamma(\tau,z)~,
\ee
with
\bea\label{thetagz}
\theta_\Gamma(\tau,z) =  \sum_{l^i\in \Gamma} e^{i\pi \tau_{ij} l_L^i\cdot l_L^j - i\pi \taub_{ij} l_R^i\cdot l_R^j+2\pi i z_{Li} \cdot l^i_L -2\pi i z_{Ri} \cdot l^i_R}~.
\eea
In this expression and in what follows we have not written the $I$ indices explicitly.
We note that, as the $U(1)^D$ descendants are uncharged, the prefactor $\Phi(\tau)$ is exactly the same as in the unflavored case.

The higher genus modular transformations act on the period matrix $\tau$ and potential $z$ as\be
\tau \to \gamma\tau\equiv (C\tau+D)^{-1}(A\tau+B)~,~~~~~z_{L,R}\to \gamma z_{L,R} \equiv (C \tau +D)^{-1} z_{L,R}~,
\ee
where $\gamma=\left({A~,B\atop C~D}\right)\in Sp(2g,\Z)$.
The salient point is that $C$ and $D$ are matrices which act on the $i=1\dots g$ indices, but not on the $I=1\dots D$ flavor indices;  therefore much of the analysis of ${\cal M}_D$ in the previous section will apply in the higher genus case as well.
The theta function transforms as
\be
\theta(\gamma\tau,\gamma z) =
\exp \left[ \pi i z_L C \left({C\tau+D}^{-1}\right) z_L - \pi i z_R C \left({C\tau+D}^{-1}\right) z_R   \right]
\theta(\tau,z)~.
\ee

\subsection{Laplace equations}

We begin, as in the genus one case, with the unflavored partition function.  The theta function is a sum over the lattice $\Gamma$ of
\bea
Q_g(l,\tau) =  e^{i\pi \tau_{ij} l_L^i\cdot l_L^j - i\pi \taub_{ij} l_R^i\cdot l_R^j}~.
\eea
As before, our goal is to write an equation relating the Laplacians on Narain moduli space and the Siegel upper half-space.

We start by writing the Siegel Laplacian.  Decomposing $\tau_{ij}$ into its real and imaginary parts as  $\tau_{ij}=x_{ij}+iy_{ij}$, the metric on Siegel upper half-space is
\bea\label{Siegmet2}
ds^2 = y^{ij}y^{kl}(dy_{ik} dy_{jl} +dx_{ik}dx_{jl})~.
\eea
Here  $y^{ij}$ denotes the inverse of $y_{ij}$, i.e $y^{ik}y_{kj}=\delta^i_j$.   It is important to note that the line element (\ref{Siegmet2}) should be expressed in terms of unconstrained variables, which we take to be $(x_{ij}, y_{ij})$ with $i\leq j$.
With this in mind, the Laplacian is
\bea\label{lap3}
 \Delta_{{\cal H}_g} = -{1\over \sqrt{g}} g^{AB} \p_A (\sqrt{g} g^{AB} \p_B) = - y_{ik}y_{jl}(\hat{\p}_{x_{ij}}\hat{\p}_{x_{kl}} +\hat{\p}_{y_{ij}}\hat{\p}_{y_{kl}})~,
\eea
where $\hat{\p}_{x_{ij}} = {1\over 2}(1+\delta_{ij}){\p \over \p x_{ij}}$ and $\hat{\p}_{y_{ij}} = {1\over 2}(1+\delta_{ij}){\p \over \p y_{ij}}$.   In (\ref{lap3}) the index sums each run from  $1$ to $g$, but $ \Delta_{{\cal H}_g}$ should be expressed in terms of the unconstrained variables.

We wish to act with the $ \Delta_{{\cal H}_g}$ on $Q_g(l,\tau)$, which we now write as
\bea\label{Qg2}
Q_g(l,\tau) = e^{i\pi x_{ij} (l_L^i \cdot l_L^j -l_R^i \cdot l_R^j )}  e^{-\pi y_{ij} (l_L^i \cdot l_L^j +l_R^i \cdot l_R^j ) }~.
\eea
A slight inconvenience is the presence of hatted derivatives in $ \Delta_{{\cal H}_g}$.   However, these can be dispensed with by the following observation.   We are instructed to express $Q_g(l,\tau)$ in terms of the unconstrained quantities $(x_{ij},y_{ij})$ with $i\leq j$ and then act with the hatted derivatives $(\hat{\p}_{x_{ij}},\hat{\p}_{y_{ij}})$.   It is simple to verify that this gives the same result as if we think of all $(x_{ij},y_{ij})$ as being independent, act with ordinary derivatives $({\p \over \p x_{ij}} , {\p \over \p y_{ij}})$, and then at the end impose $(x_{ji}=x_{ij}, y_{ji}=y_{ij})$.  This statement relies on the form $Q_g(l,\tau)$ and does not hold for all functions.
We conclude that when  acting on $Q_g(l,\tau)$ we can write
\bea\label{lap2x}
 \Delta_{{\cal}_g} = - y_{ik}y_{jl}\left( {\p \over \p x_{ij}}{\p \over \p x_{kl}} + {\p \over \p y_{ij}}{\p \over \p y_{kl}}\right)~,
\eea
and view $(x_{ji}, y_{ji})$ as independent of $(x_{ij},y_{ij})$ until the end of the computation.

This simplification in hand, we can now proceed as we did for $g=1$.
Defining
\bea\label{Tdef2x}
 T^{I}_J =\sum_{i=1}^g \left( l_L^{iI} {\p \over \p l_R^{iJ}} +  l_R^{iJ} {\p \over \p l_L^{iI}}\right)~,
\eea
the same logic as for $g=1$ leads to
\begin{align}\label{j2}
 J^2 Q_g(l,\tau)&\,  = e^{i\pi x_{ij} (l_L^i \cdot l_L^j -l_R^i \cdot l_R^j )} 2 T^I_J T^I_J  e^{-\pi y_{ij} (l_L^i \cdot l_L^j +l_R^i \cdot l_R^j )}\cr
 & \, = \left[32 \pi^2 y_{ij}y_{kl} l_L^i\cdot  l_L^k l_R^j \cdot l_R^l -8\pi D y_{ij}\left( l_L^i\cdot l_L^j+l_R^i\cdot l_R^j\right) \right] Q_g(l,\tau)~.
\end{align}
We also have the differential operator relations
\begin{align}
	 -y_{ij}y_{kl} {\p \over \p \tau_{ik}} {\p \over \p \taub_{jl}}  Q_g(l,\tau) &= -\pi^2 y_{ij}y_{kl} (l_L^i \cdot l_L^k) (l_R^j\cdot l_R^l)  Q_g(l,\tau)~, \\
	y_{ij} { \p \over \p y_{ij} }  Q_g(l,\tau) &= -\pi y_{ij} (l_L^i \cdot l_L^j +l_R^i \cdot l_R^j )~.
\end{align}
So that equation \eqref{j2} can be rewritten as
\bea
J^2  Q_g(l,\tau)  = \left[ 32 y_{ij}y_{kl} {\p \over \p \tau_{ik}} {\p \over \p \taub_{jl}} +8D y_{ij} { \p \over \p y_{ij} }\right]  Q_g(l,\tau)~.
\eea
We can write the Laplacians of the Narain moduli space and the genus-$g$ Riemann surface as
\bea
\Delta_{\cal M} = -{1\over 8}J^2~,\quad \Delta_{{\cal H}_g} = - 4 y_{ij}y_{kl} {\p \over\p  \tau_{ik} }  {\p \over\p \taub_{jl} }~.
\eea
The action of these  on $Q_g(l,\tau)$ are related in the following manner
\bea
\left[  \Delta_{{\cal H}_g}  - D  y_{ij} { \p \over \p y_{ij} } - \Delta_{\cal M} \right]  Q_g(l,\tau) =0~.
\eea
We  now use
\begin{align}\label{diffy}
\Delta_{{\cal H}_g} (\det y)^s = - {gs(2s-g-1) \over 2}  (\det y)^s~.
\end{align}
This relation is straightforward to derive by using the relations
\bea
\hat{\p}_{y_{ij}} \det y = y^{ij}~,\quad  \hat{\p}_{y_{ij}} y^{kl} = -{1\over 2} (y^{ki}y^{jl}+y^{kj}y^{jl})~,
\eea
which follow from the definitions.

We also have
\bea
\!\!\!\!\Delta_{{\cal H}_g}  \left( (\det y)^s Q_g(l,\tau) )\right) = \left[ \Delta_{{\cal H}_g} (\det y)^s \right]Q_g(l,\tau)  +  (\det y)^s  \left[ \Delta_{{\cal H}_g} Q_g(l,\tau) \right] + {\rm cross~term},
\eea
where the cross term is
\bea
-2 y_{ij} y_{kl} \left( \hat{\p}_{y_{ik}}  (\det y)^s \right) \left(\hat{\p}_{y_{jl}}Q_g(l,\tau)  \right) = -2s   (\det y)^s  y_{jl} { \p \over \p y_{jl}} Q_g(l,\tau)~.
\eea
This finally gives
\bea\label{diffe}
\left[  \Delta_{{\cal H}_g}  - \Delta_{\cal M} +  {gs(2s-g-1) \over 2}  \right] \left(  (\det y)^s Q_g(l,\tau) \right)=0~,
\eea
which is the result quoted in \cite{Maloney:2020nni}.

Summing over lattice points, we see that the theta function itself obeys the same differential equation.
As in the genus one case, we can sum equation $(\ref{diffe})$ over $Sp(2g,\Z)$ images to obtain an Eisenstein series which is a modular invariant eigenfunction of the Laplacian with the same eigenvalue.
We will write this Eisenstein series as
\bea
Y(z_L=0,z_R=0,\tau,\taub)  = (\det y)^s\sum_{(C,D)=1} {1\over |\det(C\tau+D)|^{2s}}~,
\eea
where the  notation ``$(C,D)=1$" means that the matrices $C,D$ together form the lower row of an $Sp(2g,\Z)$ matrix; of course when $g=1$ this reduces to the usual condition that $C$ and $D$ are coprime integers.\footnote{As in the $g=1$ case, this should be regarded not as a sum over $Sp(2g,\Z)$ but rather a sum over a coset $Sp(2g,\Z)/P$ where the subgroup $P$ just consists of all transformations which leave $\det y$ invariant.}

It is now straightforward to generalize this to the flavored case, since the ${\cal M}_D$ structure more or less comes along for the ride.  For example, acting on functions of $(l,z)$ the generators of the $O(D,D)$ currents now take the form
\begin{align}\label{lrt}
 T^I_J &=
 \sum_{i=1}^g
 l_L^{Ii} \cdot {\p \over \p l_R^{Ji}} +  l_R^{Ii} \cdot {\p \over \p l_L^{Ji}}+ z_{Li}^J \cdot {\p \over \p z_{Ri}^I} +  z_{Ri}^J \cdot {\p \over \p z_{Li}^I}.
\end{align}
The argument follows that given above, resulting in the final equation for the averaged partition function,
%
%
\bea
\left[ \Delta_{\cal H} +s(s-1)+{1\over 4} \left( z_{iL}^J {\p \over \p z_{Ri}^I} +  z_{Ri}^J {\p \over \p z_{Li}^I} \right)  \left( z_{Li}^J {\p \over \p z_{Ri}^I} +  z_{Ri}^J {\p \over \p z_{Li}^I} \right)  \right]\left( \det y)^{s} \langle \Theta_\Gamma(\tau,z)\rangle \right) =0~. \nonumber \\ \label{avgfdif}
\eea

\subsection{Heat equation and the Siegel-Weil formula at higher genus}

The most natural solution to the differential equation (\ref{avgfdif})  is  the Eisenstein series
\be\label{gammasol}
 \langle \Theta_\Gamma(\tau,z)\rangle
 = \sum_{(C,D)=1} {e^{ -i \pi \left( {z_L (C\tau+ D)^{-1}C z_L}\right) +i \pi \left( {z_R (C\tau +D)^{-1}C z_R}\right)} \over |\det(C\tau+D)|^{D}}~.
\ee
To demonstrate that this is indeed the correct solution, we will utilize a heat equation as in the $g=1$ case.
In particular, writing
\bea
\theta_\Gamma(\tau,z) = \sum_{l^i\in \Gamma}P(l,z,\tau)~,
\eea
with
\bea
P(l,z,\tau)=  e^{i\pi \tau_{ij} l_L^i \cdot l_L^j - i\pi \taub_{ij} l^i_R l^j_R +2\pi i z^i_L \cdot l^i_L -2\pi i z^i_R \cdot l^i_R}~,
\eea
an identical argument as described at $g=1$ gives
\bea\label{avgfdiff}
\left( {\p \over \p \tau_{ij}} -{1\over 4\pi i} {\p^2 \over \p z^i_L \cdot \p z^j_L} \right) P(l,z,\tau) =0=  \left({\p \over \p \taub_{ij}} +{1\over 4\pi i} {\p^2 \over \p z^i_R \cdot \p z^j_R}\right) P(l,z,\tau)~.
 \eea
As before, this reflects the fact that the stress tensor is Sugawara.
Summing over lattice points and integrating over Narain lattices, we find that $\theta_\Gamma(\tau,z)$ and $\langle \theta_\Gamma(\tau,z)\rangle$ both obey the differential equation  (\ref{avgfdiff}) as well.

As at genus one, this provides a relationship between the $\tau$ and $z$ dependence of the partition function which can be used to prove (\ref{gammasol}).
The important point is that one can start with the solution at $z=0$ (where (\ref{gammasol}) was proven in \cite{Maloney:2020nni}), and then expand the heat equation order by order in $z$ to develop recursive relations relating different orders in this expansion.  It is straightforward to check that  $(\ref{gammasol})$ is the unique solution to these recursion relations which obeys the correct boundary condition at $z=0$.

%
%

There is, however, one important distinction between this case and the simple $g=1$ case considered earlier.  The integral over $O(D,D)$ implies that the resulting expressions for $\langle \theta_\Gamma(\tau, z)\rangle $ will be invariant under the $O(D)\times O(D)$ symmetries which rotate the chemical potential vectors $z_{Li}^{I}$ and $z_{Ri}^I$; these rotations act on the $I=1\dots D$ indices, but not on the $i=1\dots g$ index.  The result is that the averaged partition function will be a function only of the invariants $z_L^i \cdot z_L^j$ and $z_R^i\cdot z_R^j$.  When $g=1$ this includes only the lengths of the chemical potential vectors, which we denoted $z_L^2$ and $z_R^2$.  At $g>1$, however, there are now new invariants which appear with $i\ne j$.  Loosely speaking, this reflects the fact that when the genus $g$ partition function is constructed as a sum over states (corresponding to some channel decomposition of the genus $g$ surface), charge will flow between different channels.

\subsection{A genus 2 example}

In order to illustrate the utility of (\ref{gammasol}), let us consider in more detail the genus $2$ case.\footnote{We are especially grateful to S. Collier for discussions related to the computations appearing in this subsection.}  Here the period matrix
 $\tau_{ij}=\left(\tau_1~\tau_{12}\atop\tau_{12}~\tau_2\right)$ is two dimensional, and the averaged partition function will depend as well on the inner products of the charge vectors  $z_1^2= z_{1}^I z_1^I $, $z_2^2 = z_2^I z_2^I$ and $z_1\cdot z_2=z_1^I z_2^I$ in both the left and right moving sectors.  Of particular interest is the pinching limit $\tau_{12}\to0$, where the genus two surfaces factorizes into disjoint union of two tori.  We note that, generically, the partition function of this pinched Riemann surface does not contain all of the data of our CFT.
 In the Narain case, however, as long as one keeps $z_1\cdot z_2$ non-zero, it is possible to completely reconstruct the genus two partition function in terms of the factorized torus correlators; by using the heat equation one can determine the dependence on $\tau_{12}$.    This holds at higher genus as well, and means that  if one wishes one can completely disregard the genus $g$ partition functions and instead just work with the expectation values $\langle Z(\tau_1, z_1)\dots Z(\tau_g, z_g)\rangle$ of a product of flavored partition functions; all of the dependence of the genus $g$ partition function on the moduli $\tau_{ij}$ with $i\ne j$ can be reconstructed from the factorized limit by considering the dependence on $z_i\cdot z_j$.

To understand this in more detail, we can begin by considering the averaged genus $2$ partition function
$Z(\tau, Q^{Ii},{\bar Q}^{iI})$ in a sector of fixed charge, rather than fixed potential, which is given by the Fourier transform:
\be
\langle Z(\tau, Q^{Ii},{\bar Q}^{iI})\rangle = \frac{1}{\Phi(\tau)}\int dz_L^{iI} dz_R^{iI}e^{2\pi i (z_L^{iI} Q^{iI}+z_R^{iI} {\bar Q}^{iI} )}
\langle \Theta_\Gamma(\tau,z)\rangle~,
\ee
where
\be\label{gammasola}
 \langle \Theta_\Gamma(\tau,z)\rangle
 = \sum_{(C,D)=1} {e^{ -i \pi \left( {z_L (C\tau +D)^{-1} C z_L}\right) +i \pi \left( {z_R (C\tau +D)^{-1} C z_R}\right)} \over |\det(C\tau+D)|^{2s}}~.
\ee
We will focus on the contribution of the primary states so will drop the prefactor $\frac{1}{\Phi(\tau)}$.
We will also consider the case  $\tau_{12}=0$ where this genus two partition function factorizes into a product of genus 1 surfaces.  The partition function $Z(\tau, Q^{Ii},{\bar Q}^{iI})$ can then be used to compute the two point function of the density of states:
$\langle\rho(\Delta_1,j_1, Q^{I1}) \rho(\Delta_2,j_2,Q^{I2})\rangle$.

It is possible to unpack the sum over coprime matrices $(C,D)$ following 
\cite{Maa__1971}.  This technique is applied extensively in \cite{CM}, so we will only summarize a few relevant details here.  
One starts by considering separately the cases where $C$ has rank $0$, $1$ or $2$.
The case with rank $0$ just corresponds to $(C,D)=(0,1)$, which gives 
 the contribution of the vacuum state with $Q^{iI}=0$.
When rank $C=1$, the Fourier transform of $Z(\tau, Q^{Ii},{\bar Q}^{iI})$ vanishes unless $Q^1$ is proportional to $Q^2$, and we find\footnote{
To see this, we note that the  set of coprime matrices with rank$(C)=1$ is parameterized by two pairs of coprime integers $(c,d)=1$ (with $c\ne 0$) and $(m,n)=1$, with \be
C = \left({c~0\atop 0~0}\right)U^T,~~~~~D=\left({d~1\atop 0~1}\right) U,
\ee
where $U=\left({m~p\atop n~q}\right)$ is a unimodular matrix.}   
\begin{align}
&\langle \rho(\Delta_1, j_1, Q_1^I)\rho(\Delta_2,j_2,Q_2^I)\rangle_{\rm rank\,1}
  \cr
  &  = \left[\sum_{(m,n)=1}\kappa(j_{m,n},D) \delta(mQ_1+nQ_2)\right]
\delta\left(\Delta_1-(Q_1^2+{\bar Q_1}^2)\right) \delta\left(\Delta_2-(Q_2^2+{\bar Q_2}^2)\right)\cr
& \quad\quad \times
\delta\left(j_1-(Q_1^2-{\bar Q_1}^2)\right)
\delta\left(j_2-(Q_2^2-{\bar Q_2}^2)\right)~.
\end{align}
Finally, the set of coprime matrices $(C,D)$ with rank$(C)=2$ are parameterized by symmetric matrices $P=C^{-1}D$ with rational entries, leading to
\be
\langle \Theta_\Gamma(\tau,z)\rangle_{{\rm rank}\,2} = \sum_P \nu(P)^{-2s}
{e^{ -i \pi \left( {z_L (\tau +P)^{-1} z_L}\right) +i \pi \left( {z_R (\tau +P)^{-1} z_R}\right)} \over |\det(\tau+P)|^{2s}}
\ee
where $\nu(P)=\det(C)$ is the product of the elementary divisors of $P$.
The Fourier transform gives a Gaussian integral as in the genus one case:
\begin{align}
\langle Z(\tau, Q^{Ii},{\bar Q}^{iI})\rangle_{\text{rank}\,2} &= 
e^{-2\pi (QyQ+{\bar Q}y{\bar Q})+2\pi i (QxQ-{\bar Q}x{\bar Q}) }\sum_{P} \nu(P)^{-2s} e^{2\pi i (Q P Q - {\bar Q}P {\bar Q})}   ~,\nn
\end{align}
where we have written $\tau_{ij}=x_{ij}+iy_{ij}$.  
The sum over $P$ is a version of 
Siegel's singular series\footnote{Note that $Sp(4,\Z)$ invariance implies that the partition function is invariant under $\tau\to\tau+N$ (i.e. $x\to x+N$) for any symmetric integral matrix $N$.  This implies that $QNQ-{\bar Q}N{\bar Q}$ is an integer. So we can let
$P=R+N$ where $N$ is a symmetric integral matrix and the entries of $R$ are rational numbers between $0$ and $1$, and replace $P$ by $R$ in the sum.}
\be
S_s(Q_1^2,Q_2^2,Q_1\cdot Q_2)\equiv \sum_R \nu(R)^{-2s} e^{2\pi i (Q R Q- {\bar Q} R {\bar Q})}~,
\ee 
where the sum is over rational symmetric matrices with entries between zero and one.
This series should be regarded as a generalization of the zeta function relevant for Eisenstein series of higher genus.
The Fourier transform then gives
\begin{align}
\langle \rho(\Delta_1, j_1, Q_1^I)\rho(\Delta_2,j_2,Q_2^I)\rangle_{\text{rank}\,2}
=&\,
S_s(Q_1^2,Q_2^2,Q_1\cdot Q_2)
\cr & \times
\delta\left(\Delta_1-(Q_1^2+{\bar Q_1}^2)\right)
\delta\left(\Delta_2-(Q_2^2+{\bar Q_2}^2)\right)
\cr
&\times
\delta\left(j_1-(Q_1^2-{\bar Q_1}^2)\right)
\delta\left(j_2-(Q_2^2-{\bar Q_2}^2)\right)~.
\end{align}

\section{Flavored partition function from Chern-Simons theory}
\label{sec:bulk}

In this section we show that the averaged flavored partition function can be reproduced in a natural way by summing over a class of geometries weighted by appropriate  $U(1)^D \times U(1)^D$ Chern-Simons partition functions.   As compared to the unflavored case, the new feature is that we allow for more general boundary conditions on the gauge fields, corresponding to the presence of  chemical potentials in the partition function.  These have to be treated carefully in order to respect the modular behavior of the partition.   In this section we restrict attention to the genus one CFT flavored partition function, which on the Chern-Simons side means that we restrict the class of bulk geometries to be solid tori. Our discussion is similar to \cite{Kraus:2006nb}.

We consider $U(1)^D \times U(1)^D$ Chern-Simons theory on a manifold $M$ with boundary $\partial M$.   The action is
\bea\label{SCS}
S={i\over 8\pi} \int_M \left( A^I \wedge dA^I - \Ab^I \wedge d\Ab^I\right) -{1\over 16\pi} \int_{\p M} d^2x \sqrt{g} g^{ab}\left(  A^I_a A^I_b +  \Ab^I_a \Ab^I_b\right)~,
\eea
where $g_{ab}$ is the metric on $\p M$.  The boundary term is chosen so that the on-shell variation of the action is
\bea
\delta S = {i\over 2\pi} \int_{\p M} d^2x \sqrt{g}\left(  J_I^a \delta A^I_a -  \Jb_I^a \delta \Ab^I_a \right)~,
\eea
where the currents are
\begin{align}
 J^{I}_a &= {i \over 4} \left( A^I_a-i\eps_a^{~b} A^I_b\right) , \qquad
  \Jb^{I}_a = {i \over 4} \left( A^I_a+i\eps_a^{~b} A^I_b\right)~,
\end{align}
and $(A_a^I,\Ab_a^I)$ function as their conjugate potentials.
The stress tensor is defined via the variation with respect to the metric,
\bea
\delta S = {1\over 2} \int_{\p M} d^2x \sqrt{g}\,  T^{ab}\,\delta g_{ab}~,
\eea
yielding
\bea
T_{ab} = {1\over 8\pi} \left( A^I_a A^I_b-{1\over 2} A^{Ic}A^I_c g_{ab} +  \Ab^I_a \Ab^I_b-{1\over 2} \Ab^{Ic}\Ab^I_c g_{ab} \right)~.
\eea
Choosing the flat metric $g_{ab}dx^a dx^b  = dwd\wb$, these formulas read
\begin{align}\label{JT}
&J^I_w = {i\over 2} A^I_w~,\quad J^I_{\wb} = 0~ ,\quad
\Jb^I_{w} = 0~,\quad  \Jb^I_{\wb} = {i\over 2} \Ab^I_{\wb}   ~,\\
&T_{ww} = {1\over 8\pi} \left( A^I_w A^I_w+ \Ab^I_w \Ab^I_w \right) ,~~
T_{\wb\wb} ={1\over 8\pi} \left( A^I_{\wb} A^I_{\wb}+ \Ab^I_{\wb} \Ab^I_{\wb} \right)  ,~~
T_{w\wb}  =  T_{\wb w} =0~.\nn
\end{align}
Note that the non-zero components of the stress tensor are a sum of a Sugawara piece quadratic in the currents and a contribution quadratic the potentials.

We now turn to the computation of the flavored partition function as a sum over geometries.  We let $M$ be a solid torus. We choose a radial coordinate $r$ such that at fixed $r$ we have a $T^2$ on which we choose a complex coordinate $w$.   The $w$ coordinate is taken to have periodicities
\bea
w \cong w+2\pi \cong w+2\pi \tau~,\quad \tau = \tau_1+i\tau_2~.
\eea
The boundary  cycle defined by the identification $w\cong w+2\pi$ is taken to be contractible when extended into the solid torus.

The flavored partition function is defined by fixing boundary conditions for the connection.   We fix (in this section $z_L \equiv z$, $z_R \equiv \bar z$)
\bea
A^I_{\wb} = {i\over \tau_2} z^I~,\quad \Ab^I_w =-{i\over \tau_2} \zb^I~.
\eea
Note that $z^I$ and $\zb^I$ are not related by complex conjugation.   Demanding vanishing holonomy around the contractible circle imposes
\bea
A^I_w =-A^I_{\wb}~,\quad \Ab^I_{\wb} = -A^I_w~.
\eea
For  flat connections with these boundary values, the full contribution to the classical action comes from the boundary term in (\ref{SCS}), and gives
\bea
S= -{\pi \over 2\tau_2} (z^2 + \zb^2)~,
\eea
where we are now writing $z^2= z^I z^I$ and $\zb^2 = \zb^I \zb^I$.
Since the action is quadratic, the 1-loop fluctuation determinant is not affected by the potentials $(z^I,\zb^I)$.   It is equal to the partition function of $D$ free bosons on the torus \cite{Maloney:2020nni,Porrati:2019knx}.  Altogether, the path integral for the theory on the solid torus is\footnote{The notation $(\tau,z)$ is shorthand for $(\tau,\taub, z^I,\zb^I)$.}
\bea
Z_{PI}(\tau,z) = {1\over |\eta(\tau)|^{2D}} e^{ {\pi (z^2 +\zb^2 )\over 2\tau_2}}~.
\eea
An important point is that the path integral differs from the partition function, where the latter is defined as
\bea
  Z(\tau,z)= {\rm Tr} \left[ e^{2\pi i \tau
(L_0 -c/24)} e^{-2\pi i \taub (\Lb_0-{c}/24)} e^{2\pi i z^I Q_I}e^{-2\pi i \zb^I \bar Q_I}\right]~.
\eea
In the above, $L_0$ and $\Lb_0$ take the Sugawara form, quadratic in the currents.  In the presence of chemical potentials as implemented by our boundary conditions, we noted previously that the stress tensor written in (\ref{JT}) is the sum of a Sugawara piece and a contribution from the potentials.  Taking this into account, one finds that the path integral and the partition function are not equal, but rather differ by a contribution from the potentials \cite{Kraus:2006nb},
\bea\label{Zconv}
Z(\tau,z)= e^{ -{\pi (z^2 +\zb^2 )\over 2\tau_2}}Z_{PI}(\tau,z)~.
\eea
The prefactor is responsible for the fact that while the path integral for a CFT with $U(1)$ currents is modular invariant, the partition function with nonzero potentials picks up a multiplicative factor, as written in (\ref{Zmod}).
Noting cancellation of the prefactor, the contribution to the partition function is therefore simply
\bea
Z(\tau,z) = {1\over |\eta(\tau)|^{2D}} ~.
\eea
The fact that this is independent of the potentials follows from  our assumption of trivial holonomy around the contractible cycle; this implies that no charge propagates around the non-contractible cycle.

We now include the sum over bulk manifolds, corresponding to summing over inequivalent choices for which boundary cycle is contractible in the bulk.  We can  implement this by writing $w = (c\tau+d)w'$, with identification $w'\cong w'+2\pi \cong w'+2\pi \tau'$, with $\tau'=(a\tau+b)/(c\tau+d)$.  As usual, $ad-bc=1$.  We now take the contractible cycle to be the one corresponding to the identification $w'\cong w'+2\pi$.  The classical action is given by the boundary term, which is coordinate invariant.
With $\tau'_2 = \tau_2/|c\tau+d|^2$ and
\bea
z'^I = -i\tau'_2 A^I_{\wb'} = -i {\tau_2 \over c\tau+d}A^I_{\wb} =  {z^I \over c\tau+d}~,
\eea
we obtain
\begin{align} S(\tau',z') &= -{\pi \over 2\tau'_2}(z'^2 + \zb'^2) =  -{\pi \over 2\tau_2}(z^2 + \zb^2) + \pi i \left( {cz^2 \over c\tau+d}-{c\zb^2 \over c\taub+d}\right) ~.
\end{align}
Using $|\eta(\tau')|^2 = |c\tau+d||\eta(\tau)|^2$, we find that the contribution to the path integral is
\bea
Z^{(c,d)}_{PI}(\tau,z) ={ e^{ {\pi (z^2+\zb^2)\over 2\tau_2}}\over |\eta(\tau)|^{2D}}    { e^{- \pi i \left( {cz^2 \over c\tau+d}-{c\zb^2 \over c\taub+d}\right) } \over |c\tau+d|^D}~.
\eea
We convert the partition function using (\ref{Zconv}) and, following \cite{Maloney:2007ud}, sum over inequivalent geometries labelled by relatively prime integers $c$ and $d$ to get
\begin{align}
Z(\tau,z) &= \sum_{(c,d)=1}e^{ -{\pi (z^2 +\zb^2 )\over 2\tau_2}} Z^{(c,d)}_{PI}(\tau,z)  =      {1 \over |\eta(\tau)|^{2D}}     \sum_{(c,d)=1}   { e^{- \pi i \left( {cz^2 \over c\tau+d}-{c\zb^2 \over c\taub+d}\right) } \over |c\tau+d|^D}~.
\end{align}
This reproduces our previous expression (\ref{ZFavg})  for  the averaged flavored partition function.

As in the unflavored case, we can think of extending this computation to higher genus boundaries. The classical action will again come from boundary terms, with boundary conditions that fix the holonomy around all  boundary cycles that are non-contractible in the bulk. This classical part will reproduce terms in the flavored Siegel-Narain theta function (\ref{thetagz}).    The one-loop contribution, denoted as $1/\Phi(\tau)$, is much more complicated than  at genus one, but we again expect it to be independent of the boundary conditions since the action is quadratic.

\section*{Acknowledgements}
We are grateful to S. Collier, A. Dymarsky, K. Jensen, A. Shapere and E. Witten for useful conversations.
P.K. is supported in part by the National Science Foundation under research grant PHY-1914412.
Research of AM is supported in part by the Simons Foundation Grant No.~385602 and the Natural Sciences and Engineering Research Council of Canada (NSERC), funding reference number SAPIN/00032-2015.

\begin{small}
\providecommand{\href}[2]{#2}\end{small}

\end{document}